\documentclass[12pt]{iopart}
\usepackage{graphicx}

\begin{document}
\title{Folksonomies and clustering in the collaborative system CiteULike}
%\shorttitle{Taxonomy and clustering in Wikipedia} %Insert here a short version of the title if it exceeds 70 characters

\author{A Capocci$^1$ and G Caldarelli$^2$}

\address{$^1$ Dip. di Informatica e Sistemistica
  Universit\`a ``Sapienza'', via Ariosto, 25 00185 Rome, Italy}
\address{$^2$ SMC Centre, INFM-CNR, Dip. di Fisica, Universit\`a
  ``Sapienza'', P.le A. Moro 2, 00185-Rome, Italy}

\begin{abstract}
We analyze {\em CiteULike}, an online collaborative tagging system
where users bookmark and annotate scientific papers. Such a system can
be naturally represented as a tripartite graph whose nodes represent
papers, users and tags connected by individual tag assignments. The
semantics of tags is studied here, in order to uncover the hidden
relationships between tags. We find that the clustering coefficient
reflects the semantical patterns among tags, providing useful ideas
for the designing of more efficient methods of data classification and
spam detection..
\end{abstract}


\maketitle

The recent development of the World Wide Web is characterized by a
growing number of online social communities. In many such cases,
individuals provide bits of information - about either their tastes,
opinions or interests - and software applications gather and organize
them into a database, allowing the browsing of the whole information
collected so. A class of such collaborative systems focusses on
collecting users' online bookmarks with either a general approach
or a more specialized one. In particular, some websites have been
recently born to store user--generated scientific bibliographies.

In these systems, the elementary contribution, the ``post'', is made
of three ingredients: a user, an article and an annotation of it by a
number of tags chosen by users. In exchange for this voluntary
contribution, a user can browse others' bibliographies and
annotations. Tags are an alternative classification method with
respect to traditional taxonomies, where items belong to ``taxa''
represented as a tree--like set of categories: here, each category
contains in turn a number of more specialized sub--categories, and so
on until the desired resolution of classification is been
reached. Instead, in tagging systems items are tagged by users
characterized by diverse tagging strategies depending on a number of
individual variables. The set of tag--resource relations in such a
community is called a ``folksonomy''.

Such communities are now extremely popular, storing hundreds of
thousands posts and more. The tagging system we analyze here, {\em
CiteULike} \cite{citeulike}, has been built, at the time of our
survey, by ca. 180000 references annotated by ca. 48000 tags supplied
by ca. 6000 users. Our dataset includes about 550000 ``tag
assignments'': each assignment is a t-uple (user, resource, tag). The
sequence of chronologically ordered tags, in particular, can be
interpreted as a stream of words, to which one can applies the
traditional statistical text analysis to uncover how human behavior
affects it.

The statistical analysis of word occurrences in a written text has
shown that word frequencies are power--law distributed according to
the Zipf's law, according to which a large number of words appears in
a text only a few times, while a few words occur orders of magnitude
more often \cite{zipf49}. Such feature has been modeled by many models
based on the preferential attachment principle, that is, the
assumption that authors employ already used words with a probability
proportional to the current word frequency. Moreover, it has been
observed that the rate of new words decrease with the text--length
\cite{sublineargrowth}, that is, the number of distinct word $N_w$ in
a text of length $L$ scale as
\begin{equation}
N_w \propto L^\beta,
\end{equation}
with $\beta<1$. However, models in literature assume that new words
are introduced at a constant growth rate, so that their total number,
i.e. the vocabulary, is a linear constant of the total number of words
(both new or repeated ones) used so far
\cite{yule25,simon55,ferrer05}.

Yet, to discover the semantical properties of {\em CiteULike}, one rather
represents it by means of the network formalism, which proved fruitful
in the analysis of many natural and social phenomena involving
unsupervised interacting units: in a network perspective, elementary
interacting agents or objects are represented by nodes, interactions
by edges connecting them. The widespread success of such approach has
been triggered by the discovery that many networks instances one
encounters in reality share common statistical properties with no
external tuning. For example, the degree $k$, that is, the number of
edges pointing to a node, follows in many cases a broad distribution
$P(k)$ with long tail decaying algebraically as 
\begin{equation}
P(k) \propto k^{-\gamma}, 
\end{equation}
with $\gamma<3$. If edges have varying intensities, each
of them is attached a weight $w$ representing its intensity;
accordingly, a node is characterized by its $strength$, equal to the
sum of the weights of edges pointing to it. The distribution of
$weights$, too, is power--law distributed in many real weighted
network instances. Furthermore many such networks exhibit a strong
transitivity, i.e. with high probability, the neighbors of a node are
themselves connected by an edge, with respect to purely random
realization of a network with equal number of nodes and
links. Networks sharing the above properties are currently named
``complex networks'' \cite{caldarellibook}.

The network approach has also been recently adopted to analyze the
semantical structure of tagging systems
\cite{huberman,lambiotte06,ripeanu}. Tags can be
represented by networks in different ways, in order to study how the
behavior of users maps into the dynamical or topological features. A
more recent stream of research deals with the organization of tags,
which are implicitly linked by hierarchical and logical associations
emerging despite the diversity of users as their number is large
enough. The underlying semantical organization of tags reveals the
dominant trends within a tagging community and allows to improve its
navigability. Recently, algorithms have been introduced in order to
infer a taxonomy of tags from a folksonomy
\cite{garciamolina07,stumme07}.

The statistical properties we observed in the {\em CiteULike} data are
consistent with the findings obtained in similar surveys, confirming
that tags in collaborative systems form complex networks
indeed. Moreover, we have investigated how the underlying semantics of
tags reflects on the topology of the network.

As a matter of facts, tags provided by users come with no explicit
hierarchy beside the chronological ordering, leading authors to
analyze the stream of tags as a text--like sequence of
words. Interestingly, the time--ordered sequence of tags displays
statistical properties already observed in written texts, such as the
fat--tails in the word frequency distributions or the sub-linear
vocabulary growth.

\begin{figure}
\centering
\includegraphics[width=0.8\textwidth]{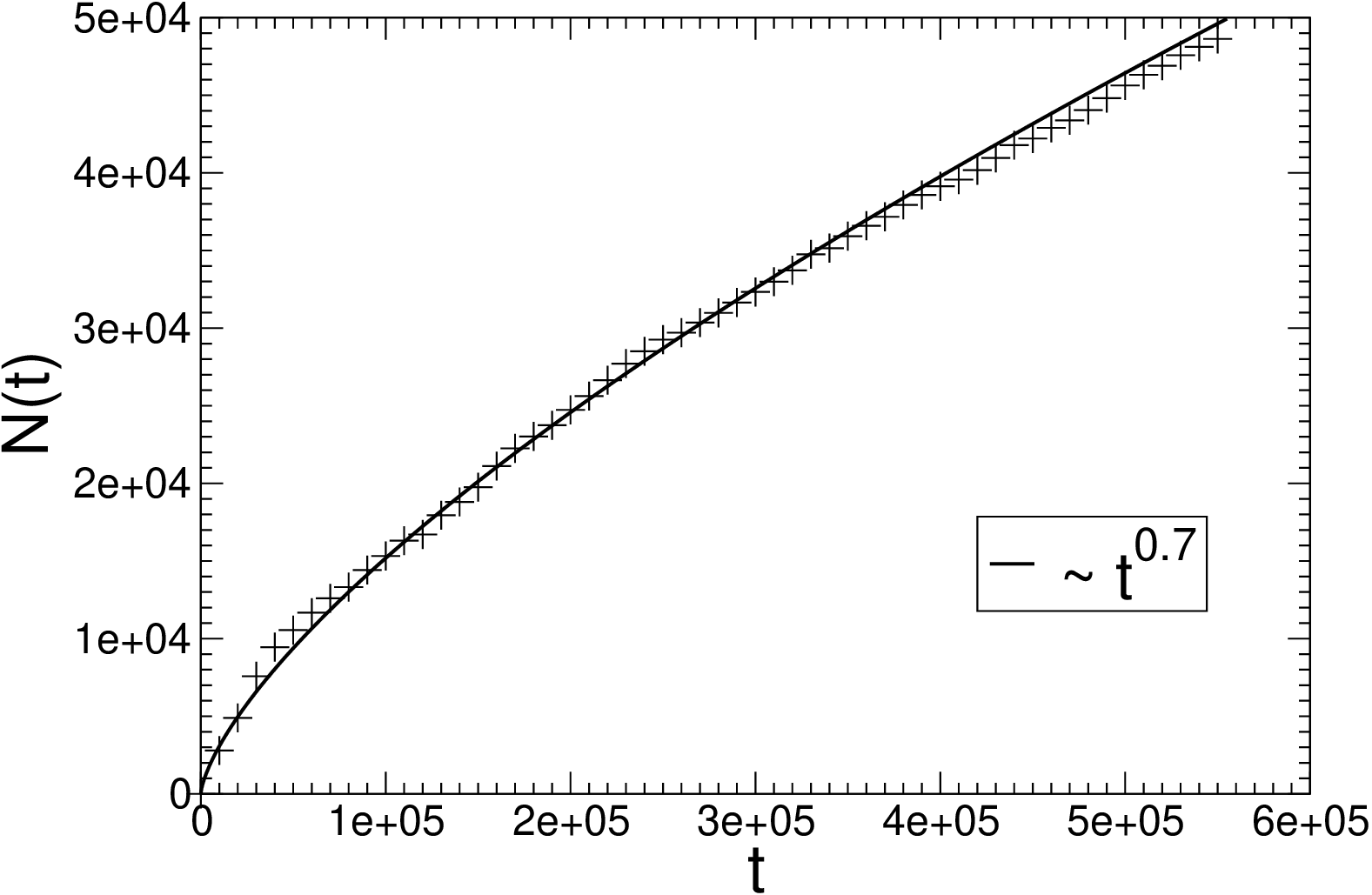}
\caption{ The number of tags $N(t)$ as a function of time $t$, where
time is measure in chronologically ordered tags assignments (plus
symbols). Solid line represents $t^{0.7}$ for a comparison.}
\label{sublinear}
\end{figure}

Our analysis confirms the sub-linear vocabulary growth observed in
written texts. The number of distinct tags $N(t)$ introduced by users
after $t$ assignments grows approximately as $N(t) \propto t^{0.7}$,
as shown in figure \ref{sublinear} although the pace is slightly
smaller than in other collaborative tagging systems already surveyed
\cite{cattuto06}. 

\begin{figure}
\centering
\includegraphics[width=0.8\textwidth]{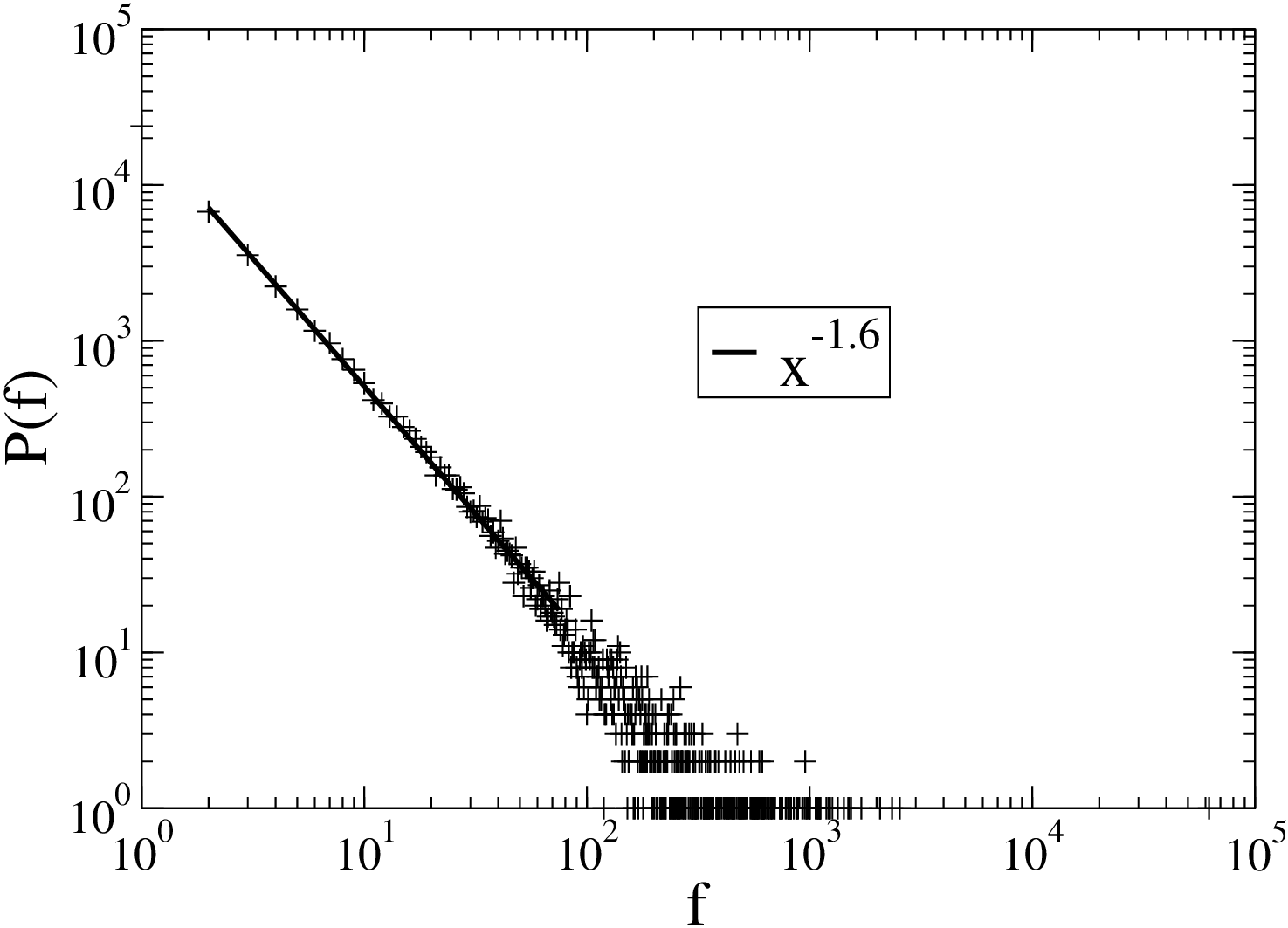}
\caption{ The statistical distribution $P(f)$ of tag frequencies $f$
(plus symbols). Solid line represents $f^{-1.6}$ for a comparison.}
\label{wordsfreq}
\end{figure}

The frequency of tags, too, reported in \ref{wordsfreq}, reminds that
of words observed in written texts, algebraically decaying according
to the Zipf's law \cite{zipf49}.

%Furthermore, text--writing models often rely on the preferential
%attachment rule, that is, the assumption that already used words are
%employed again with a probability increasing proportionally with the
%current word frequency. Such a pattern is present indeed in the growth
%of the {\em CiteULike} vocabulary, at least for small values of the
%frequency as shown in \ref{prefatt}.

However, the sole frequency of tags as a function of time does not
convey much information about the semantics, although it reflects the
different centrality of associated concepts in the underlying
knowledge organization. To investigate tag pair relations, one has to
represent the unit elements of a tagging system as nodes of a network.

The dataset we focus on can be naturally represented as a tri--partite
network, where each node represents either a user $u$, a resource $r$
or a tag $t$; if a tag assignment $(u,r,t)$ exists, an edge is drawn
from $u$ to $r$, and from $r$ to $t$. Since in a single user's post a
resource can be tagged more than once, one post can correspond to
multiple tag assignments \cite{lambiotte06}.

Although efficient algorithm have been developed to analyzed such
tri--partite network \cite{tagora-folkrank}, the heterogeneity of
nodes discourages in general the application of traditional network
methods, mainly conceived to deal with network connections
representing peer-to-peer relationships. Thus, to study how tags are
organized we chose to project the tri--partite networks on the tag
space. As a result, the tag co--occurrence network we study is
composed by nodes representing tags only, between which an undirected
edge of weight $w$ is drawn if $w$ distinct resources are labeled by
both tags.

\begin{figure}
\centering
\includegraphics[width=0.8\textwidth]{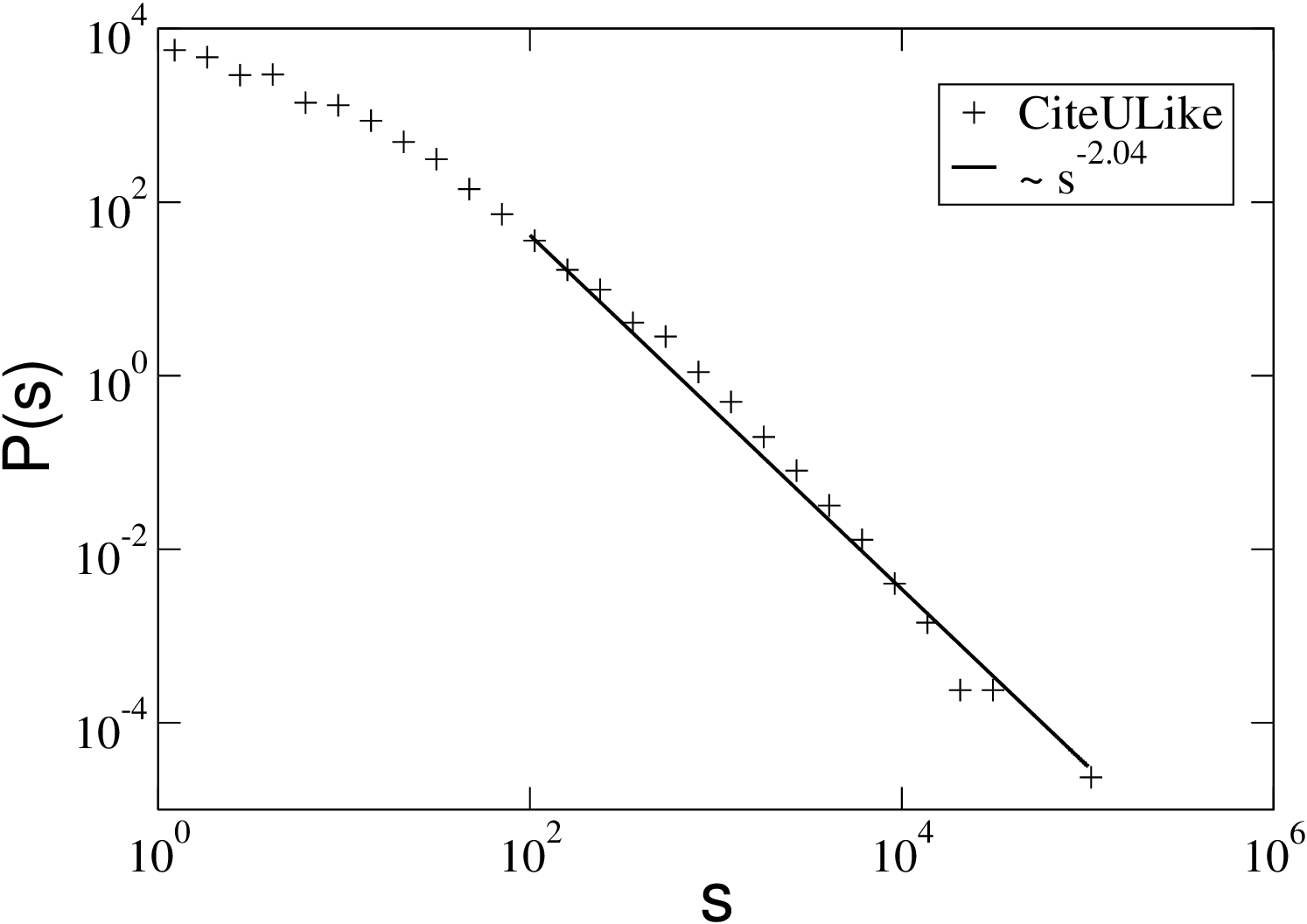}
\caption{ 
The distribution $P(s)$ of the node strengths $s$ in the tag
co--occurrence network. The best fitting power--law exponent,
represented by the solid curve, yields $\gamma = 2.04$.
}
\label{strength-dist}
\end{figure}

The resulting network displays some of the typical features of
weighted scale--free networks. We have measured the distribution of
the sum $s$ of the weights of edges pointing to a given node, or the
{\em strength} of the node: such distribution $P(s)$, plotted in
figure \ref{strength-dist}, exhibits a clear power--law decay $P(s)
\propto s^{-\gamma}$, with $\gamma = 2.04 \pm 0.02$ for large values of
$s$. Interestingly, the heterogeneity of the observed nodes' weights
does not necessarily reflects the centrality of corresponding
concepts, that are supposed to be assigned together with a wider range
of more specialized concepts, in the underlying hierarchy of tags. As
it has been already shown \cite{cattuto_knn}, reshuffling the tag
assignment in order to destroy the logical association among words
does not change dramatically the shape of $P(s)$, which proves that
the $weight$ heterogeneity is more a consequence of frequency
distribution broadness than of the varying roles of concepts in the
semantical organization of the whole vocabulary.
 
Nevertheless, the tag co--occurrence network unveils some semantical
feature of the underlying ontology if, instead of focussing on the
properties of single nodes, one turns to the inspection of quantities
involving its environment. An example of such is represented by the
analysis of the neighbor average degree $K_{nn}(k)$ of nodes with
degree $k$, where the degree is the number of incoming edges of a
node.

In our study we examined instead the clustering properties of the tag
co--occurrence network through the clustering coefficient. Such
coefficient $C(k)$ counts the average density of triangles involving
nodes with degree $k$ or, in other words, the probability that the
nearest neighbors of a node with degree $k$ are in turn connected one
to each other. This reads

\begin{equation}
C(k) = \frac{2\sum_{i, k(i)=k}\sum_{j>h}^{1,k} a_{ji}a_{hi}}{N_k k(k-1)},
\end{equation}

where $a_{ij}$ is 1 if a link exists between $i$ and $j$ and 0
otherwise, and $N_k$ is the frequency of nodes with degree $k$. This
quantity has been found to characterize most complex networks found in
nature and society, where it takes substantially larger values with
respect to a purely random networks \cite{barabasi-review}. The
properties of the clustering coefficient are often associated to the
hierarchical organization of nodes \cite{clustering-hierarchical}.

\begin{figure}
\centering
\includegraphics[width=0.8\textwidth]{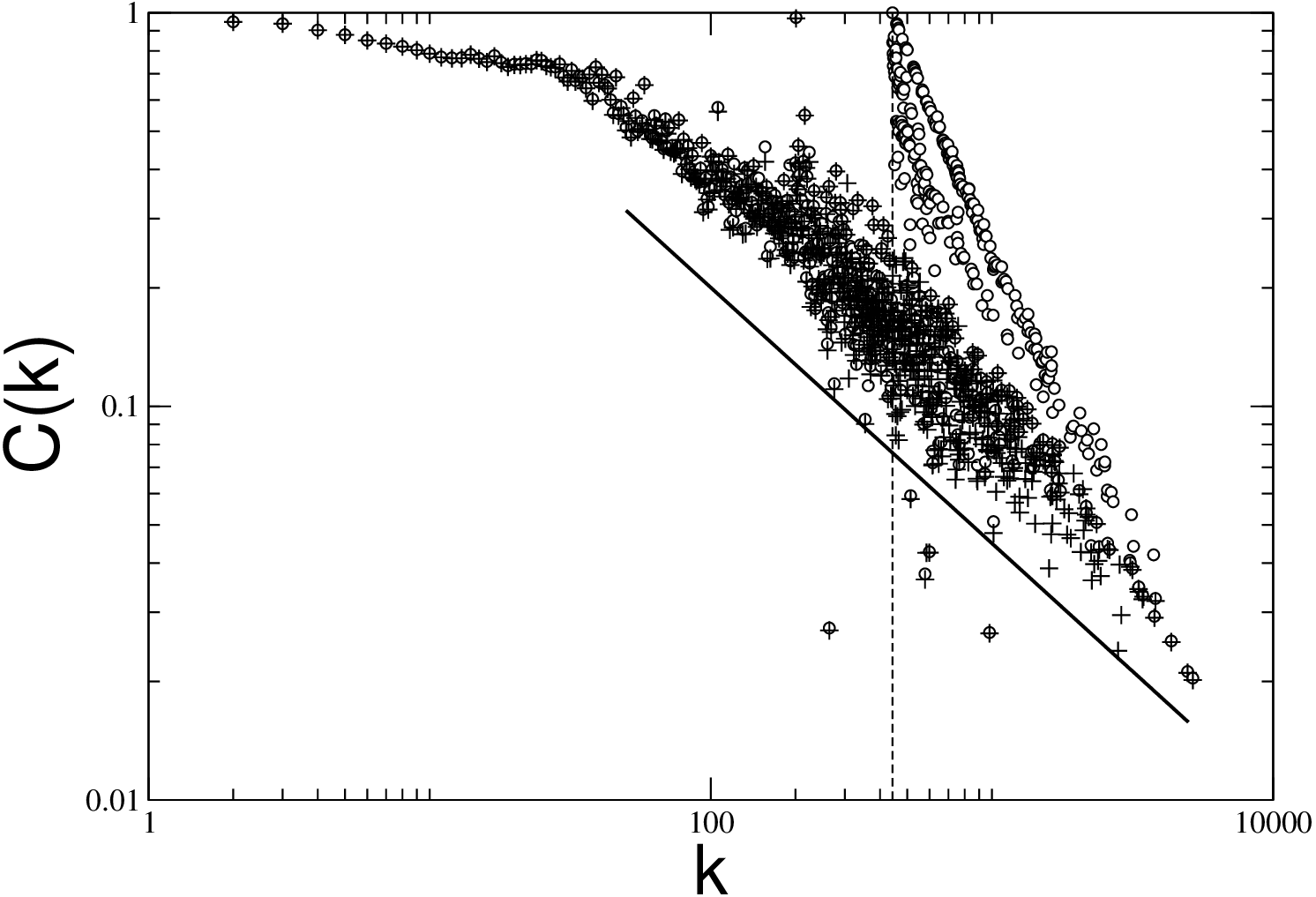}
\caption{ The clustering coefficient $C(k)$ of the tag co--occurrence
network as a function of the nodes' degree $k$ before (circles) and
after (plus symbols) the removal of a spam post from the dataset. The
solid line represents a decay $k^{-0.64}$ and the dashed vertical
ruler is set at $k=443$.}
\label{clustering-real}
\end{figure}

Indeed, the clustering coefficient appears to encode a signature of
semantical relations between words. As represented in figure
\ref{clustering-real}, the clustering coefficient $C(k)$ in {\em CiteULike}
decays algebraically for large values of the degree $k$, according to
$C(k) \propto k^{-0.64}$. However, the clustering value displays an
apparent fluctuation at $k=443$. By inspecting the nodes corresponding
to such value, one discovers that the sharp rise taking place at al
$k=443$ corresponds to a non--existing resource labelled by $444$
distinct uncorrelated randomly chosen tags, which mimics the a spam
contribution to the collaborative systems.

One is led thus to conjecture that the overall semantical organization
of concept represented by tags is encoded in a characteristic behavior
of the clustering coefficient $C(k)$, so that tags assigned in a
semantically inconsistent way fall far away from this behavior. To
verify such conjecture, we have performed the same statistical
analysis after removing from the data set the tag assignments related
to the spam--like page. As shown in figure \ref{clustering-real},
after the removal the clustering coefficient follows a more regular
behavior, confirming that the strong fluctuation observed above was
due indeed to the presence of a single meaningless set of assignments
involving a single resource. Tags assigned only to the spam resource
form a complete co--occurrence network, so that their clustering
coefficient is equal to $1$.

Thus, the behavior of clustering coefficient of the tag co--occurrence
networks can be used as a test for models representing the tag
semantical organization or, equivalently, how users choose tags when
annotating a resource. As noted in literature \cite{garciamolina07},
users typically use tags hierarchically, labelling a resource by tags
related to the same topics but with different generality, adding more
specialized tags as the number of collected resources grows.

On a very basic level, we have tested how such hierarchical tagging,
affects the topology of the tag co--occurrence network by a simple toy
model defined in the following. Let us assume that tags are organized
on a taxonomy, that is, a tree--like structure stemming from a seed
node, where each node corresponds to a tag and is an offspring of
another tag belonging to the same branch of knowledge with higher
generality. At discrete time steps, a new post is added to the system,
with a new resource and 2 tags. The first tag can be a new one, with
probability $p_g$: in such case, the new tag is an offspring of a tag
randomly chosen among the already employed ones. Otherwise, the first
tag is chosen at random among the already employed ones. The second
tag is either chosen at random from within the whole set of used tags
or, with probability $p_b$, it is chosen according to hierarchy: in
such case, the second tag is drawn at random among the nodes that lie
on the shortest path length from the first tag to the seed node on the
tree--like taxonomy.

The tag co--occurrence network resulting from the above algorithm
share some features of the {\em CiteULike} one, if we assume a
time--dependent $p_g$ which reproduces the sub-linear vocabulary growth
observed in reality, and by a suitable choice of the parameter $p_b$,
which mimics the relevance of hierarchy in tagging activity. To
reproduce the growth rule, we have set
\begin{equation}
p_g(t) = A t^{-B}
\end{equation}
and imposed that 
\begin{equation}
\int_{1}^{N_{res}} p_g(t) dt = N_{tag}
\end{equation}
and 
\begin{equation}
N_{tag} = N_{ass}^{\beta},
\end{equation}
where the number of resources $N_{res}$, the number of tag assignments
$N_{ass}$, the number of tags $N_{tag}$ and $\beta$ are set to the same values
they take in {\em CiteULike}. As a result, this yields $A=5$ and $B=0.3$.

\begin{figure}
\centering
\includegraphics[width=0.8\textwidth]{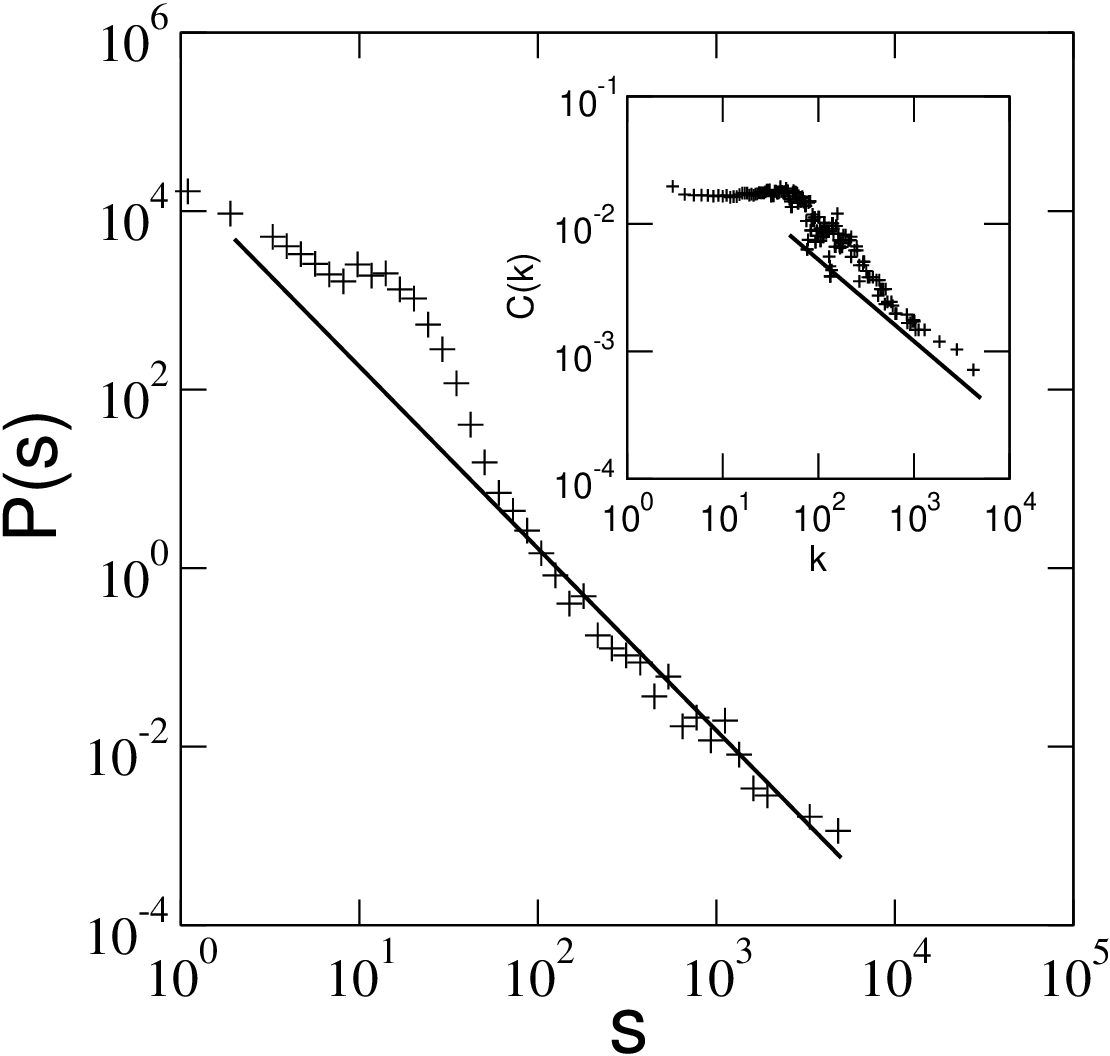}
\caption{ Main plot: the strength distribution in the co--occurrence
network derived from the model with $p_b = 0.25$ (plus symbols); the
solid line represents the decay $s^{-2.04}$ for a comparison with real
data. Inset: the clustering coefficient in the co--occurrence network
derived from the model with $p_b = 0.25$ (plus symbols). The solid
line represents the decay $k^{-0.64}$ for a comparison with real
data.}
\label{summarymodel}
\end{figure}

As shown in figure \ref{strength-dist}, the strength
distribution $P(s)$ of tags is a scale--free one with a good agreement
with reality in the decaying exponent for large values of $s$ if one
sets $p_b=0.25$. For such choice of the parameter, the clustering
coefficient reproduces qualitatively the algebraic decay observed in
{\em CiteULike}, as shown in figure \ref{summarymodel}, although the
absolute value differs of orders of magnitude.

We have found, thus, a simple model that captures the complex features
of a tag co--occurrence network issued from the dataset describing an
online collaborative tagging system, {\em CiteULike}. In particular,
by assuming that users label resources by hierarchically associated
tags, the probability distribution of nodes' strength is reproduced
for a suitable choice of the parameters. Moreover, the model
reproduces qualitatively the decaying asymptotic behavior of the
clustering coefficient $C(k)$. Such quantity encodes a signature of
the semantical organization of concepts represented by tags, so that
malicious or meaningless tag assignments can be detected by inspecting
the perturbation to the clustering coefficient they
generate. Establishing a relationship between clustering and semantics
may suggest tools and algorithms for technological tasks such as
automatic categorization of resources, recommendation and spam
detection techniques.

%\begin{figure}
%\includegraphics[width=7cm]{prefatt.eps}
%\caption{
%}
%\label{prefatt}
%\end{figure}

The authors acknowledge useful discussions with Francesca Colaiori,
Stefano Leonardi, Ciro Cattuto, Vito D.P. Servedio and Andrea
Baldassarri. The authors acknowledge the European project DELIS for
support and R. Cameron for providing the data.

\end{document}